\documentclass[preprint]{aastex631}

\usepackage{amsmath}
\allowdisplaybreaks
\usepackage{xcolor}
\submitjournal{ApJ}

\shorttitle{Analytical modelling}
\shortauthors{Matcheva et al.}
\graphicspath{{./}{figures/}}

\begin{document}

\title{Analytical Modelling of Exoplanet Transit Specroscopy with Dimensional Analysis and Symbolic Regression}

\correspondingauthor{Katia Matcheva}
\email{matcheva@ufl.edu}

\author[0000-0003-4182-9096]{Konstantin T. Matchev}
\affiliation{Physics Department, University of Florida, Gainesville, FL 32653, USA}

\author[0000-0003-3074-998X]{Katia Matcheva}
\affiliation{Physics Department, University of Florida, Gainesville, FL 32653, USA}

\author[0000-0003-2719-221X]{Alexander Roman}
\affiliation{Physics Department, University of Florida, Gainesville, FL 32653, USA}



\begin{abstract}
The physical characteristics and atmospheric chemical composition of newly discovered exoplanets are often inferred from their transit spectra 
which are obtained from complex numerical models of radiative transfer.
Alternatively, simple analytical expressions provide insightful physical intuition into the relevant atmospheric processes. The deep learning revolution has opened the door for deriving such analytical results directly with a computer algorithm fitting to the data. As a proof of concept, we successfully demonstrate the use of symbolic regression on synthetic data for the transit radii of generic hot Jupiter exoplanets to derive a corresponding analytical formula. As a preprocessing step, we use dimensional analysis to identify the relevant dimensionless combinations of variables and reduce the number of independent inputs, which improves the performance of the symbolic regression. The dimensional analysis also allowed us to mathematically derive and properly parametrize the most general family of degeneracies among the input atmospheric parameters which affect the characterization of an exoplanet atmosphere through transit spectroscopy.

\end{abstract}



\tableofcontents

\newpage 

\section{Introduction} 
\label{sec:intro}

Over the last two and a half decades the number of known planets outside the Solar System has soared from only a handful to several thousand\footnote{The Extrasolar Planet Encyclopedia, \url{http://exoplanet.eu}.} and the focus of the science community has shifted from planet detection to statistical studies and characterisation of the environments of these distant worlds \citep{Madhusudhan2019}. A number of current and planned large-scale planetary surveys are based on observing {\em planetary transits} at different wavelengths. During a transit event, a planet blocks a certain fraction, $M(\lambda)$,  of the original stellar flux $F_O(\lambda)$,
\begin{equation}
M(\lambda) \equiv \frac{F_O(\lambda) - F_T(\lambda)}{F_O(\lambda)},
\label{eq:Mdef}
\end{equation}
where $F_T(\lambda)$ is (the minimum of) the observed flux during transit at a given wavelength $\lambda$.

The transit spectroscopy targets the detection of commonly present atmospheric gasses that have strong absorption lines in the infrared and leave a distinct imprint on the observed modulation, $M(\lambda)$, of the stellar flux \citep{Schneider1994,Charbonneau2000,Seager2000}. The theoretical basis of transit spectroscopy has been developed and discussed in a number of studies \citep{Brown2001,Hubbard2001,Burrows2003,Fortney2005,Benneke2012,deWit2013,Griffith2014,Vahidinia2014,Heng2015,Betremieux2017,Heng2017} and has been successfully used to extract information about the temperature, composition and cloud opacity of the atmospheres of numerous transiting exoplanets \citep{Cobb2019,Barstow2020,Kitzmann2020,Harrington2021,Cubillos2021,Blecic2021,Welbanks2021}. There are several radiative transfer models that perform detailed calculations of the absorption of the stellar flux as it is being attenuated by the gas surrounding the planet. These models typically incorporate collisionally induced absorption (CIA) provided by the main atmospheric gasses, line-by-line calculation of the absorption coefficients of minor gas components,  and wave-independent cloud opacity based on the grey cloud approximation. Atmospheric refraction and scattering are believed to have a higher order effect on the observed flux and are mostly excluded from the simulations \citep{Seager2000,Brown2001,Hubbard2001}. The complexity of the underlying atmosphere can vary a great deal: from a one-dimensional, isothermal, well-mixed atmosphere (the temperature, the gas mixing ratios and the cloud opacity are fixed to a constant value with no altitude, latitude or longitude variations) to three-dimensional models that have variable temperature profile and vertically resolved cloud layers. Some models even explore the day-night asymmetry in the properties of the atmosphere as the leading and the trailing hemispheres are probed at sunset or sunrise during the transit \citep{MacDonald2021}.

The information content of the recorded transit spectrum, $M(\lambda)$, has been a focus of discussion of several studies \citep{Griffith2014,Heng2017,Welbanks2019,Welbanks2021atmospheric}. It has been pointed out that the parameters and/or structure of the underlying atmosphere cannot be uniquely determined from these observations without additional independent information. \cite{Heng2017} used a simple analytical approach to derive a formula for the observed effective radius of a transiting planet as a function of the atmospheric structure and chemical composition. Despite the simplicity of their formulation, the derived analytical result nicely illustrated the limitations of transit spectroscopy to uniquely determine all atmospheric parameters. 

Ultimately, the primary goal of exoplanet transit spectroscopy is the inversion of the observed spectrum in order to retrieve the parameters of the planet and its atmosphere. This process inevitably relies on a forward model which can produce large data sets of synthetic spectra, whose dependence on the underlying atmospheric parameters, however, is often obscured by the model complexity. This is why it is of great interest to have relatively simple analytical expressions as substitutes for the complicated (and slow) forward model. This not only provides valuable insights into the relevant atmospheric processes, but also helps guide the thought process during the inversion.

Recently machine learning (ML) is increasingly being used in the analysis of spectroscopic data from exoplanet transits \citep{Marquez2018,Cobb2019,Fisher2020,Guzman2020,Nixon2020,Yip2021}. Most of the time it is used to solve the inverse problem, i.e., to do the retrieval of the parameter values given the spectroscopic observations. Here we shall focus on applications of ML to the forward modelling itself. There are two possible approaches.

\begin{itemize}
\item {\em Numerical approach.} Replace the slow, complex and accurate full-blown simulation of the forward model with a fast, simple and approximate deep-learning model \citep{Himes2020Marge}. The deep-learning model is typically trained on data simulated by the existing forward model. The advantages of this approach are: i) once it is trained, the model offers significant speedup; ii) it opens the door to non-experts to participate without the need to know all the specifics of the full-blown forward model. However, there are also certain disadvantages: i) the deep-learning model is in principle a black box which hides the relevant physics \citep{Yip2021}; ii) the deep-learning model learns not the physics of the forward model itself, but the (finite amount of) data generated by the forward model, and this introduces additional uncertainties due to the training process \citep{Matchev:2020tbw}.
\item {\em Analytical approach.} Here one asks the machine to derive an analytical formula which describes the data well \citep{Schmidt2009,Udrescu:2019mnk}. In order to succeed, the search for analytical formulas needs to i)  use the relevant (combinations of) variables and ii) avoid irrelevant (combinations of) variables. How to properly accomplish all of these tasks within the analytical approach is the main focus of this paper.
\end{itemize}

One of the two major goals of this paper is to demonstrate the use of symbolic regression to derive accurate analytical formulas representative of a typical forward model. For concreteness, we shall use the recently proposed PySR framework \citep{Cranmer:2020wew} which leverages the advantages of both deep learning (graph neural networks) and symbolic regression. Similar studies have been done in several physical domains \citep{Battaglia2016,Chang2016,Udrescu:2019mnk,Iten2020,Arechiga2021}, but to the best of our knowledge, not in exoplanetary science.

A typical forward model takes as inputs a relatively large number of input parameters (on the order of a dozen or more). Keeping all of them as independent degrees of freedom significantly complicates the task of symbolic regression. Fortunately, one can use dimensional analysis to restrict the relevant number of degrees of freedom. The second major goal of this paper is to demonstrate the use of dimensional analysis to identify the relevant combinations of physical parameters (the so-called Pi groups \cite{Barenblatt1996}) which uniquely determine the observed transit spectra. In particular, we shall show that the initial set of seven free parameters in our example reduces to only four potentially relevant degrees of freedom. This simplification has important benefits:
\begin{itemize}
\item The reduction in the relevant degrees of freedom greatly enhances the performance of the symbolic regression.
\item Our dimensional analysis correctly reproduces the parameter degeneracies already known in the literature \citep{Heng2017, Welbanks2019} and identifies many new ones.
\item The main result from our dimensional analysis not only agrees with previously derived analytical approximations in the literature \citep{Heng2017}, but also points to the only allowed extensions of those existing formulas which are consistent with basic physics.
\end{itemize}

The paper is organized as follows. In Section~\ref{sec:model} we review the pros and cons of the two different approaches to the forward problem of transit spectroscopy and motivate our course of action in this paper. In Section~\ref{sec:DimensionalAnalysis} we apply dimensional analysis to the radiative transfer problem: we first derive the set of dimensionless variables describing the problem and then in Section~\ref{sec:test} identify the relevant ones among them. As a byproduct, in Sections~\ref{sec:DimensionalAnalysis} and \ref{sec:test} we derive the most general set of degeneracies that arise among the atmospheric parameters. In Section~\ref{sec:regression} we use our parametrization in terms of dimensionless variables to fit a symbolic regression and obtain analytic expressions for the forward radiative transfer model. In Appendix~\ref{sec:appendix} we compile a list of simple degeneracies among the atmospheric parameters.

\section{Modelling the Physics of Radiative Transfer in the Atmosphere} 
\label{sec:model}

\subsection{Detailed forward numerical simulations}

\begin{figure}[t]
\centering
\includegraphics[width=.55\textwidth]{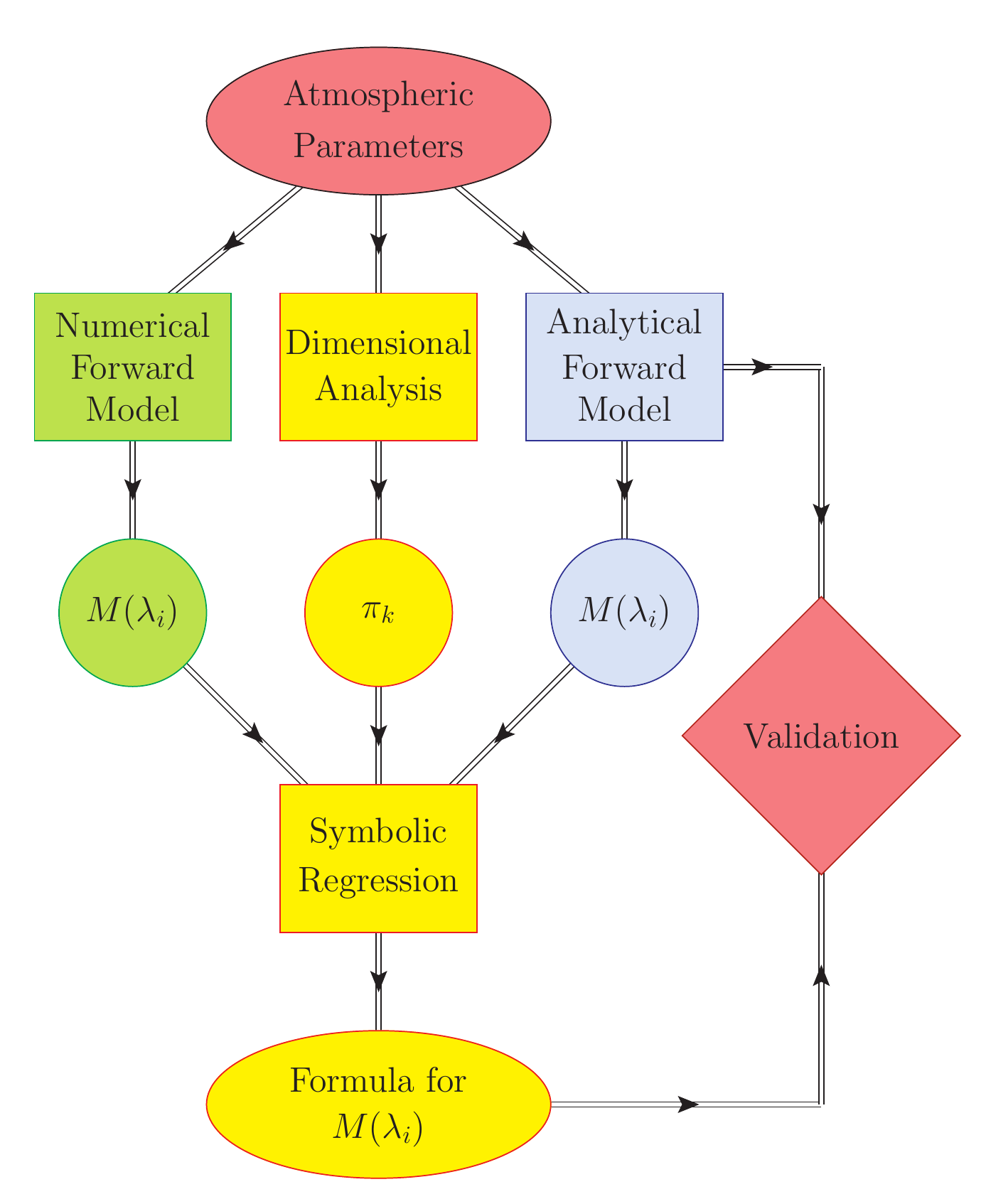}
\caption{A simplified flowchart illustrating the numerical (left green-shaded branch) and the analytical (right blue-shaded branch) alternatives for a forward radiative transfer model. The central yellow-shaded branch represents the dimensional analysis/symbolic regression approach in this paper.
\label{fig:flowchart}}
\end{figure}

At the heart of all observationally driven methods used to derive the properties of a planetary atmosphere is a numerical forward radiative transfer model, schematically depicted with the green-shaded rectangle in the left branch of Fig.~\ref{fig:flowchart}. The model starts with a given set, $\mathcal{S}$, of atmospheric parameters: temperature ($T$) and pressure ($P$) profiles, chemical abundances ($n_j$) of the present gases, mean molecular mass ($m$), cloud opacity ($\kappa_{cl}$), specific gravity ($g$), and geometry (reference planet radius $R_0$ and stellar radius $R_S$). The model then calculates the atmospheric transmission of the planet and generates a synthetic spectrum of the emerging specific stellar flux, $F_T(\lambda_i)$, at several different wavelengths $\lambda_i$: 
\begin{equation}
    \mathcal{S}(T,P,n_j,m,\kappa_{cl},g, R_0, R_S)\longrightarrow F_T(\lambda_i),
\label{eq:forward}    
\end{equation}
which can then be converted to respective modulations $M(\lambda_i)$ of the observed stellar flux according to eq.~(\ref{eq:Mdef}). For an excellent short summary and historical perspective of the published exoplanetary radiative transfer models please see \cite{Blecic2021}. The forward models allow for the construction of an arbitrary complex atmosphere by implementing varying temperature profiles and chemical abundances, localized cloud layers, and can even introduce non-equilibrium processes by coupling with thermochemical models \citep{Matcheva2005,Giles2015,Changeat2019}. A more realistic description of the atmosphere, however, leads to increased complexity of the models, which necessarily translates to increased computational cost. Moreover, due to the limited information content of the observations, the increased complexity of the forward model does not necessarily equate to a more robust characterization of the atmosphere. The reliability of the derived results and the interpretation of unresolved correlations between different  atmospheric parameters are a main focus of development \citep{Heng2017,Welbanks2019}.   

\subsection{Simplified analytical approximations}

As shown in the right (blue shaded) branch of Fig.~\ref{fig:flowchart}, a viable alternative to the detailed computational radiative transfer models is the use of analytical expressions which provide physical insight into how the underlying atmospheric structure and composition directly impact the observed spectral flux.  The analytical expressions are easy to implement and understand, they are well suited for investigating correlations between different parameters and the model uncertainties are well understood and easily calculated. The trade-off is that the analytical expressions are based on simplifying assumptions, for example: isothermal atmosphere, grey clouds, spherical symmetry, etc. At the same time, such approximations seem appropriate at this early stage of exoplanet exploration, since the limited information content in the observed spectrum does not allow one to probe the complexity of the atmosphere.

In this paper we shall proceed along the central (yellow-shaded) branch of Fig.~\ref{fig:flowchart}, which uses symbolic regression to derive an analytical expression for the modulation $M(\lambda_i)$ of the observed stellar flux. The inputs to the symbolic regression procedure (represented with the lower yellow-shaded rectangle in Fig.~\ref{fig:flowchart}) are two different types of data:
\begin{itemize}
    \item {\em The input atmospheric variables or combinations of them.} Instead of feeding all input variables directly into the symbolic regression, we first perform dimensional analysis in order to reduce the number of degrees of freedom by identifying the relevant dimensionless variables $\pi_k$. 
    \item {\em The modulation of the stellar flux.} As shown in Fig.~\ref{fig:flowchart}, the predicted $M(\lambda_i)$ which enters the symbolic regression can be taken either from a numerical forward model (left branch), or from its analytical counterpart (right branch). For definiteness, the analytical model we employ here (i.e., the blue-shaded rectangle in Fig.~\ref{fig:flowchart}) uses the expression derived in \cite{Heng2017}.
\end{itemize}

Note that the left and right branches in Fig.~\ref{fig:flowchart} represent the paradigm of traditional computing, whereby given a numerical input, a program produces a numerical output; while the central branch in Fig.~\ref{fig:flowchart} represents the paradigm of machine learning, whereby given numerical inputs and the corresponding outputs, the program produces a method, in this case a symbolic formula.

Finally, the result from the symbolic regression can be compared to existing analytical expressions for forward models (the red diamond in Fig.~\ref{fig:flowchart}). This validation process can go in both directions: 
\begin{itemize}
    \item In the case when the input $M(\lambda_i)$ to the symbolic regression is taken from the right (analytical) branch, the derived formula should be identical to the one used in the forward model. This provides an important consistency check and validation of the symbolic regression procedure, which will be performed in Section~\ref{sec:regression} below.
    \item Conversely, in the case when the input $M(\lambda_i)$ to the symbolic regression is taken from the left (numerical) branch, the derived formula from symbolic regression should in principle be an improvement over the approximate analytical expressions used in the forward analytical models. This type of exercise, however, is beyond the scope of this paper and will be undertaken in a future study.
\end{itemize}

\section{Dimensional Analysis}
\label{sec:DimensionalAnalysis}

The first step in modelling any physical phenomenon is the identification of the relevant physics variables and then finding a relation among them which describes the process of interest. For sufficiently simple systems, such a quantitative relationship can be obtained from first principles, using the known fundamental physics laws. However, for sufficiently complex phenomena like the radiative transfer within an exoplanet atmosphere, such {\em ab initio} theory is often difficult, if not impossible, and one has to resort to alternative, typically numerical, modelling methods. For example, the recent wide spread of machine learning methods has led to attempts to replace the forward model with a deep-learned model (or a variant thereof) \citep{Marquez2018, Zingales2018, Cobb2019, Oreshenko2020, Himes2020, Himes2020proc, Ardevol2021}. At the same time, a simple alternative to the numerical modelling approaches is the tried and true method of dimensional analysis, which relies on the simple fact that physical laws do not depend on the choice of the basic units of measurements. In this section we shall therefore apply dimensional analysis to our problem at hand, namely, relating the observed transit radius $R_T(\lambda)$ to the relevant variables characterizing the atmosphere of an exoplanet. We shall not give a formal introduction to the method of dimensional analysis, for which we refer the interested reader to the classic books on the subject \citep{Langhaar1951,Barenblatt1996}, instead, we shall use the worked out example below to illustrate how the method works.

\subsection{Choice of variables}

The first step, the choice of the so-called {\em governing variables} in the parlance of dimensional analysis, is easy --- the parameters which might impact the {\em governed variable} $R_T(\lambda)$ were already mentioned in Section~\ref{sec:model} (see eq.~(\ref{eq:forward})). For convenience, they are collected in Table~\ref{tab:variables}, together with their names, notation and SI units. For consistency, we follow the notation of \cite{Heng2017}. We assume an isothermal atmosphere with a mean molecular mass $m$ and temperature $T$. The atmospheric structure is defined with respect to a reference pressure level $P_0$ at a given radius $R_0$, which corresponds to an optically thick atmosphere along the line of sight. The chemical composition, the gas mixing ratios, and the cloud opacity are incorporated in the gas absorption cross-section per unit mass, $\kappa$.   

\begin{table}
	\centering
	\caption{Notation, names and SI units for the physical governing variables impacting the transit radius $R_T$. The last column lists the power with which the respective variable enters the defining equation (\ref{eq:powers}).}
	\label{tab:variables}
	\begin{tabular}{lccr} 
		\hline
		Notation & Name & SI unit & Power\\
		\hline
		$R_0$ & reference radius & $m$ & $\alpha$\\
		$R_S$ & stellar radius & $m$ & $\beta$\\
		$k_B$   & Boltzmann constant & $m^2\, kg\, s^{-2}\, K^{-1}$ & $\gamma$\\
		$T$ & temperature & $K$ & $\delta$\\
		$m$ & mean molecular mass & $kg$ & $\varepsilon$\\
		$g$ & surface gravity & $m\, s^{-2}$ & $\zeta$\\
		$P_0$ & reference pressure at $R_0$ & $m^{-1}\, kg\, s^{-2}$ & $\eta$\\		
		$\kappa$ & cross-section per unit mass & $m^2\, kg^{-1}$ & $\theta$\\			\hline
	\end{tabular}
\end{table}

We notice that there are two variables, the reference radius $R_0$ and the star radius $R_S$, which are commensurable with (i.e., have the same units as) the transit radius $R_T(\lambda)$. While either one of them can be chosen to provide the dimensions of $R_T(\lambda)$, for definiteness and without loss of generality, we shall choose $R_0$, since it is more directly related to the transit radius.

\subsection{Identification of the complete set of Pi groups}

With those preliminaries, we can now write down the desired relationship in the form 
\begin{equation}
R_T(\lambda) \sim R_0 \times \left[\, R_0^\alpha\,  R_S^\beta\, k_B^\gamma\, T^\delta\, 
m^\varepsilon\, g^\zeta\, P_0^\eta\, \left(\kappa(\lambda)\right)^\theta\, \right],
\label{eq:powers}
\end{equation}
where the quantity within the square brackets is dimensionless. Dimensional homogeneity \citep{Barenblatt1996} implies that the powers $\alpha$, $\beta$, $\gamma$, \ldots, $\theta$ can be chosen to be integers. Note that there are 8 governing variables and 4 SI units appearing in Table~\ref{tab:variables}; it is easy to check that the rank of the dimensional matrix is indeed $8-4=4$. In this case, the Buckingham $\pi$ theorem \citep{Buckingham1914} then states that the quantity within the square brackets on the right-hand side of (\ref{eq:powers}) is a function of exactly four dimensionless parameter combinations $\pi_k$ with $k=1,2,3,4$ (the so-called Pi groups) constructed from the governing variables (see Fig.~\ref{fig:flowchart}). To find those, we need to solve the linear equations 
\begin{subequations}
\begin{eqnarray}
\alpha + \beta + 2 \gamma + \zeta - \eta + 2\theta \,&=&\, 0, \label{eq:unitm}\\
\gamma + \varepsilon + \eta - \theta \,&=&\, 0, \label{eq:unitkg} \\
-2\gamma - 2 \zeta -2 \eta \,&=&\, 0, \label{eq:unitsec}\\ 
- \gamma + \delta \,&=&\, 0, \label{eq:unitK}
\end{eqnarray}
\label{eq:units}
\end{subequations}
which ensure the correct dimensionality of eq.~(\ref{eq:powers}). Eliminating $\delta$, $\eta$ and $\theta$ from eqs.~(\ref{eq:unitkg}-\ref{eq:unitK}), we rewrite (\ref{eq:unitm}) in the form
\begin{equation}
\alpha + \beta + 3\gamma + 2\varepsilon + 0\zeta = 0.
\label{eq:dotproduct}
\end{equation}
This equation describes a four-dimensional flat hyperplane in the five-dimensional space spanned by the coordinates $(\alpha, \beta , \gamma , \varepsilon , \zeta )$. Eq.~(\ref{eq:dotproduct}) defines this hyperplane as being orthogonal to the constant vector $\vec{\pi}_0\equiv(1,1,3,2,0)$, therefore we can choose to parametrize points on the hyperplane as linear combinations of the following four\footnote{Of course, any other choice of four linearly independent vectors orthogonal to $\vec{\pi}_0$ will work as well. We shall take advantage of this freedom below to simplify the obtained Pi groups.} vectors which are i) orthogonal to $\vec{\pi}_0$ and ii) mutually orthogonal among themselves, which ensures that they are also linearly independent:
\begin{subequations}
\begin{eqnarray}
\vec{\pi}_1: (\alpha, \beta , \gamma , \varepsilon , \zeta ) = (1, 1, -2, 2, 0) ~~ &\Rightarrow&~~ (\delta,\eta,\theta) = (-2,2,2), \nonumber\\
\vec{\pi}_2: (\alpha, \beta , \gamma , \varepsilon , \zeta ) = (0, 0, 0, 0, -1) ~~&\Rightarrow&~~ (\delta,\eta,\theta) = (0,1,1), \nonumber\\
\vec{\pi}_3: (\alpha, \beta , \gamma , \varepsilon , \zeta ) = (1, -1, 0, 0, 0) ~~&\Rightarrow&~~ (\delta,\eta,\theta) = (0,0,0), \nonumber \\
\vec{\pi}_4: (\alpha, \beta , \gamma , \varepsilon , \zeta ) = (1, 1, 0, -1, 0) ~~&\Rightarrow&~~ (\delta,\eta,\theta) = (0,0,-1). \nonumber 
\end{eqnarray}
\nonumber
\end{subequations}
Each of these vectors in turn defines a dimensionless Pi group of governing variables:
\begin{subequations}
\begin{eqnarray}
\vec{\pi}_1 ~&\Longrightarrow&~ \pi_1 = \frac{R_0\, R_S\, m^2 P_0^2 \kappa^2}{k_B^2T^2}, \label{eq:pi1}\\
\vec{\pi}_2 ~&\Longrightarrow&~ \pi_2 = \frac{P_0 \kappa}{g}, \label{eq:pi2} \\
\vec{\pi}_3 ~&\Longrightarrow&~ \pi_3 = \frac{R_0}{R_S}, \label{eq:pi3} \\
\vec{\pi}_4 ~&\Longrightarrow&~ \pi_4 = \frac{R_0R_S}{ m \kappa}. \label{eq:pi4}
\end{eqnarray}
\end{subequations}
Note that while the first Pi group looks complicated, it can be rewritten as
\begin{equation}
\frac{R_0\,R_S\, m^2 P_0^2\, \kappa^2}{k_B^2\,T^2} =
\left(\frac{R_0\, m g}{k_B\,T}\right)^2
\left(\frac{P_0\, \kappa}{g}\right)^2
\frac{R_S}{R_0}= 
\left(\frac{R_0\, m g}{k_B\,T}\right)^2 \frac{\pi_2^2}{\pi_3} 
\nonumber
\end{equation}
and can therefore be traded for the much simpler dimensionless combination
\begin{equation}
\pi'_1 = \frac{R_0\, m g}{k_B\,T} \equiv \frac{R_0}{H},
\end{equation}
where $H$ is the pressure scale height 
\begin{equation}
H \equiv \frac{k_B\, T}{m\, g}.
\label{eq:Hdef}
\end{equation}
We can similarly trade $\pi_4$ in (\ref{eq:pi4}) for 
\begin{equation}
\pi'_4=\frac{R_0^2}{m\, \kappa}.
\end{equation}

The final result from our dimensional analysis is that any meaningful expression for the transit radius must be of the form
%
\begin{equation}
R_T(\lambda) \sim R_0 \times
f\left(\pi'_1,\pi_2(\lambda),\pi_3,\pi'_4(\lambda)\right)
= R_0 \times
f\left(\frac{R_0}{H},\frac{P_0\, \kappa(\lambda)}{g},\frac{R_0}{R_S},\frac{R_0^2}{ m \kappa(\lambda)}\right),
\label{eq:dimanalysis}
\end{equation}
where $f$ is an unspecified function of dimensionless quantities only, i.e., of the four Pi groups listed as its arguments, plus possibly some numerical dimensionless mathematical constants like real numbers, $\pi$, $e$, etc. Equation~(\ref{eq:dimanalysis}) can be recast in terms of the modulation $M(\lambda_i)=(\pi R_T^2(\lambda))/(\pi R_S^2)$ of the observed stellar flux as follows:
\begin{equation}
M(\lambda) \sim
 \left(\frac{R_0}{R_S}\right)^2 \times
f^2\left(\pi'_1,\pi_2(\lambda),\pi_3,\pi'_4(\lambda)\right)
= \left(\frac{R_0}{R_S}\right)^2 \times
f^2\left(\frac{R_0}{H},\frac{P_0\, \kappa(\lambda)}{g},\frac{R_0}{R_S},\frac{R_0^2}{ m \kappa(\lambda)}\right).
\label{eq:modulation}
\end{equation}

Equations~(\ref{eq:dimanalysis}) and (\ref{eq:modulation}) are the main results from our dimensional analysis. At this stage, the functional dependence $f$ remains arbitrary, and to make further progress, one has to fit those equations to data generated by a forward model (either numerical or analytical). In particular, the data will reveal i) whether {\em all four} $\pi$ variables are entering the actual relationship; ii)  what is the exact functional form of $f$. We shall undertake these two exercises in the next two sections, respectively. But before we do that, we conclude this section with a discussion of the parameter degeneracies implied by eqs.~(\ref{eq:dimanalysis}) and (\ref{eq:modulation}).

\subsection{Guaranteed degeneracies in the interpretation of exoplanet transmission spectra}
\label{sec:guaranteed}

The ability of transmission spectra to uniquely constrain the atmospheric parameters of exoplanets has been an intense topic of discussion\footnote{For a recent review and a guide to the relevant literature, see \cite{Welbanks2019} and references therein.} since the inception of the field. Many previous studies have noticed various degeneracies between different sets of atmospheric parameters, which result in the same observed transit spectra (within the error bars). The methods used in those works range from numerical to statistical to semi-analytical. The results from our dimensional analysis here, eqs.~(\ref{eq:dimanalysis}) and (\ref{eq:modulation}), can now help us understand from first principles the existing statements in the literature concerning the degeneracy of atmospheric retrievals. We note that the degeneracies which we discuss here are {\em exact}, and cannot be lifted by improving the experimental precision.

\begin{table}
	\centering
	\caption{Explicit parametrization of the three degrees of freedom degeneracy discussed in the text. The three parameters were chosen to be the scaling factors for $R_0$, $P_0$ and $\kappa$, respectively.}
	\label{tab:degeneracies}
	\begin{tabular}{cccc} 
		\hline
		Variable & $R_0$ scaling & $P_0$ scaling & $\kappa$ scaling\\
		\hline
		$R_0$ & $L_{R_0}$ & --- & ---\\
		$P_0$ & --- & $L_{P_0}$ & ---\\		
		$\kappa$ & --- & --- & $L_{\kappa}$ 			\\ \hline
		$T$ & $L^3_{R_0}$ & $L_{P_0}$  & --- \\
		$m$ & $L^2_{R_0}$ & --- & $L_{\kappa}^{-1}$\\
		$g$ & --- & $L_{P_0}$ & $L_{\kappa}$ \\
		$R_S$ & $L_{R_0}$ & --- & ---\\
		\hline
	\end{tabular}
\end{table}

The main point is that there are 7 variable parameters\footnote{Not counting the fundamental constant $k_B$ which presumably has the same value on the exoplanet \citep{Duff:2014}.} in Table~\ref{tab:variables}, while the physics of the transmission spectra depends on only 4 unique Pi groups (see eqs.~(\ref{eq:dimanalysis}) and (\ref{eq:modulation})). This implies that there exists a degeneracy among all seven atmospheric variables which i) can be parametrized by $7-4=3$ degrees of freedom and ii) preserves the values of {\em all four} $\pi_k$ variables, and consequently, the values of $R_T(\lambda)$ and $M(\lambda)$. We shall choose $R_0$, $P_0$ and $\kappa$ as our free-varying variables that scale as $R_0\to L_{R_0} \times R_0$, $P_0\to L_{P_0} \times P_0$ and $\kappa\to L_{\kappa} \times \kappa$, where $(L_{R_0}, L_{P_0}, L_{\kappa})$ are the three scaling factors parametrizing the degeneracy. Then, in order to keep all four $\pi_k$ groups constant, the remaining four atmospheric variables, $T$, $m$, $g$ and $R_S$, should scale as shown in Table~\ref{tab:degeneracies}. Explicitly:
\begin{subequations}
\begin{eqnarray}
R_0 ~~&\to&~~ L_{R_0} \times R_0 \\
P_0 ~~&\to&~~ L_{P_0} \times P_0 \\
\kappa ~~&\to&~~ L_{\kappa} \times \kappa \\
T ~~&\to&~~ L^3_{R_0} L_{P_0} \times T \\
m ~~&\to&~~ L^2_{R_0} L_{\kappa}^{-1} \times m \\
g ~~&\to&~~ L_{P_0} L_{\kappa}  \times g \\
R_S ~~&\to&~~ L_{R_0} \times R_S 
\end{eqnarray}
\label{eq:degeneracyscalings}
\end{subequations}
Equation~(\ref{eq:degeneracyscalings}) is the most general transformation for which the degeneracy is {\em guaranteed} in the sense that its existence does not depend on the functional form of $f(\pi'_1,\pi_2,\pi_3,\pi'_4)$. Even broader degeneracies may arise if the function $f$ does not explicitly depend on one or more of its $\pi_k$ arguments. In analogy to classical Lagrangian mechanics, we shall refer to such missing variables as ``cyclic" variables. In the next section, we shall find that this does actually occur and that $\pi'_4$ is cyclic, which further enlarges the set of degeneracy transformations. For now we conclude this section by listing the three individual degeneracies from Table~\ref{tab:degeneracies} and eq.~(\ref{eq:degeneracyscalings}):
\begin{subequations}
\begin{eqnarray}
L_{R_0}\ne 1, L_{P_0}=L_{\kappa}=1 ~~&\Longrightarrow&~~ R_0 \to L \times R_0,~~ T \to L^3 \times T, ~~ m \to L^2 \times m, ~~ R_S \to L \times R_S; \label{eq:LR0}\\
L_{P_0}\ne 1, L_{R_0}=L_{\kappa}=1 ~~&\Longrightarrow&~~ P_0 \to L \times P_0,\quad T \to L \times T, \quad g \to L \times g; \label{eq:LP0} \\
L_{\kappa}\ne 1, L_{P_0}=L_{R_0}=1 ~~&\Longrightarrow&~~ \kappa \to L \times \kappa,\quad m \to L^{-1} \times m, \quad g \to L \times g. \label{eq:Lkappa} 
\end{eqnarray}
\label{eq:Lscalings}
\end{subequations}
The first reflects a planet radius-stellar radius-temperature-mass degeneracy; the second reveals a pressure-temperature-gravity degeneracy, while the last one implies a mass-gravity-opacity degeneracy. Of course, one can also consider arbitrary {\em combinations of} those three individual transformations, as shown in eq.~(\ref{eq:degeneracyscalings}).

Each one of the transformations (\ref{eq:Lscalings}) has a clear physics interpretation. Consider, for example, eq.~(\ref{eq:LP0}). A higher temperature $T$ would make the atmosphere ``puff up", but this can be compensated by a corresponding increase in the specific gravity $g$, so that the scale height $H$ remains constant. As a result, all three length variables $R_0$, $R_S$ and $H$ remain the same, which in turn keeps $\pi'_1$ and $\pi_3$ constant as well. Furthermore, since $m$ and $\kappa$ are unaffected by the transformation (\ref{eq:LP0}), $\pi'_4=R_0^2/(m\kappa)$ is also constant. Finally, since the pressure $P_0$ and $g$ are varied at the same rate, $\pi_2\sim P_0/g$ stays constant as well.

Note that so far we have been considering the transit radius $R_T(\lambda)$ measured at a single value of $\lambda$, but all of our previous conclusions also apply for multi-wavelength observations.

\section{Identifying Relevant and Irrelevant Pi Groups}
\label{sec:test}

The dimensional analysis performed in the previous section identified the four relevant variables $\pi'_1$, $\pi_2$, $\pi_3$ and $\pi'_4$, but did not guarantee that the function $f$ in eq.~(\ref{eq:dimanalysis}) depends on all four of them. In this section we shall use three different approaches to investigate in more detail the dependence of $f(\pi'_1, \pi_2, \pi_3, \pi'_4)$ on each of its four arguments.

\subsection{Comparison to existing analytical approximations}
\label{sec:comparison}

We begin with an easy shortcut --- simply looking up the expected functional form of $f(\pi'_1, \pi_2, \pi_3, \pi'_4)$ in the existing analytical approximations in the literature. For example, \cite{Heng2017} derived the following expression for the transit radius
\begin{equation}
R_T(\lambda) = R_0 \biggl\{1 + \frac{H}{R_0} 
\biggl[\gamma_E + E_1(\tau_0) 
+ \ln (\tau_0 )
\biggr]\biggr\},
\label{eq:fullformula}
\end{equation}
where $\gamma_E=0.577215665$ is the Euler–Mascheroni constant,
\begin{equation}
\tau_0(\lambda) \equiv \frac{P_0\kappa(\lambda)}{g}\sqrt{2\pi \frac{R_0}{H}}
\end{equation}
is the optical thickness of the atmosphere along the line of sight at the reference radius $R_0$ and
\begin{equation}
E_1(\tau_0) = \int_{\tau_0}^\infty \frac{e^{-t}}{t} dt
\end{equation}
is the exponential integral of the first order with argument $\tau_0$.

In the large $\tau_0$ limit the $E_1$ term vanishes:
$$
\lim_{\tau_0\to \infty} E_1(\tau_0) = 0
$$
and (\ref{eq:fullformula}) simplifies to
\begin{equation}
R_T(\lambda) = R_0 \biggl\{1 + \frac{H}{R_0} 
\biggl[\gamma_E + \ln \biggl(\frac{P_0\kappa(\lambda)}{g}\sqrt{2\pi \frac{R_0}{H}}\biggr)
\biggr]\biggr\}.
\label{eq:simpleformula}
\end{equation}

Both eqs.~(\ref{eq:fullformula}) and (\ref{eq:simpleformula}) are consistent with our general result (\ref{eq:dimanalysis}) derived earlier. In particular, (\ref{eq:fullformula}) allows us to identify the function $f$ as 
\begin{equation}
f (\pi'_1,\pi_2) = 1 + \frac{1}{\pi'_1} \left[\gamma_E + E_1\left(\pi_2\sqrt{2\pi \pi'_1}\right) + \ln \left(\pi_2\sqrt{2\pi \pi'_1}\right)\right] ,
\label{eq:fpi1pi2full}
\end{equation}
where one should not confuse the usual constant $\pi=3.14159265$ with the four Pi groups $\pi_k$ derived in Section~\ref{sec:DimensionalAnalysis}. In the large $\tau_0$ limit, this reduces to a corresponding analogue of (\ref{eq:simpleformula}):
\begin{equation}
f (\pi'_1,\pi_2) = 1 + \frac{1}{\pi'_1} \left[\gamma_E  + \ln \left(\pi_2\sqrt{2\pi \pi'_1}\right)\right] .
\label{eq:fpi1pi2approx}
\end{equation}
The analogous expressions for the modulation $M(\lambda)$ of the observed stellar flux are:
\begin{equation}
M(\lambda; \pi'_1,\pi_2,\pi_3)= \pi_3^2 \left\{1 + \frac{1}{\pi'_1} \left[\gamma_E + E_1\left(\pi_2\sqrt{2\pi \pi'_1}\right) + \ln \left(\pi_2\sqrt{2\pi \pi'_1}\right)\right] \right\}^2
\label{eq:Mpi1pi2full}
\end{equation}
for the general case, and
\begin{equation}
M(\lambda; \pi'_1,\pi_2,\pi_3)= \pi_3^2 \left\{1 + \frac{1}{\pi'_1} \left[\gamma_E  + \ln \left(\pi_2\sqrt{2\pi \pi'_1}\right)\right] \right\}^2
\label{eq:Mpi1pi2approx}
\end{equation}
in the large $\tau_0$ limit.

Notice that the function $f$ depends on only two of the identified $\pi_k$ variables ($\pi'_1$ and $\pi_2$), while the modulation $M(\lambda)$ depends on only three out of the four variables ($\pi'_1$, $\pi_2$ and $\pi_3$). In each case, $\pi'_4$ is missing and can be identified as a cyclic variable, thus expanding the previously discussed set of degeneracies (\ref{eq:degeneracyscalings}).

In what follows we shall illustrate our results with the synthetic benchmark data set of \cite{Marquez2018} which consists of 100,000 synthetic Hubble Space Telescope Wide Field Camera 3 (WFC3) spectra of transit radii of hot Jupiters observed at 13 different wavelengths $\lambda_i$ in the range $0.838-1.666\, {\mu}m$. The data set is created by using equation (\ref{eq:simpleformula}) and scanning the parameter space of five different atmospheric parameters: temperature $T$, abundances of the H${}_2$O, NH${}_3$, and HCN gases, and grey cloud opacity $\kappa_{cl}$.

\begin{figure*}[t]
\centering
\includegraphics[width=.6\textwidth]{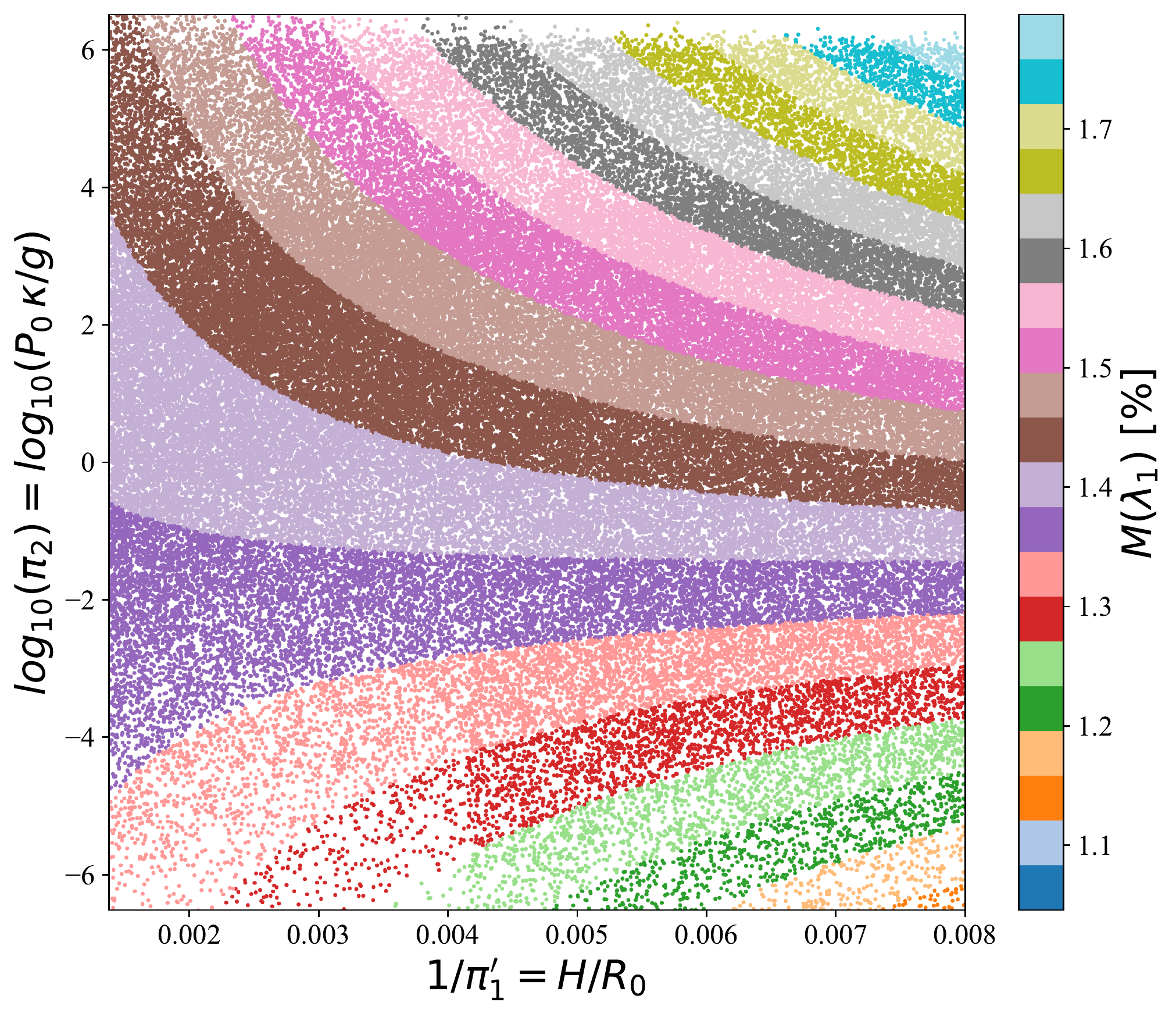}
\caption{Scatter plot of the 100,000 points in the data set of \cite{Marquez2018}  versus the dimensionless parameters $1/\pi'_1$ and $\log_{10}(\pi_2)$, color coded by the value of $M(\lambda_1)$ in percent.
\label{fig:scatter}}
\end{figure*}

Figure~\ref{fig:scatter} shows a scatter plot of the 100,000 points in the data set of \cite{Marquez2018}  versus our dimensionless parameters $\pi'_1$ and $\pi_2$, where for plotting convenience, we have chosen to display $1/\pi'_1$ and $\log_{10}(\pi_2)$ on the axes. The points are color coded by the value of $M(\lambda_1)$ in percent, where $\lambda_1 = 0.867\, {\mu}m$ is the first wavelength in the data set. Note that $R_0$ and $R_S$ were kept fixed when producing the data set, and therefore $\pi_3$ does not vary throughout it, which prevents us from using this data to illustrate any $\pi_3$ dependence.

The well-defined color bands in Figure~\ref{fig:scatter} indicate that $M(\lambda_1)$ is a unique function of the two plotted variables, $\pi'_1$ and $\pi_2$, as suggested by the generating formula in eq.~(\ref{eq:simpleformula}). This confirms that the relevant physics information encoded in the spectroscopic measurements throughout this synthetic data set is indeed only sensitive to $\pi'_1$ and $\pi_4$ (since $\pi_3$ was fixed), in agreement with eq.~(\ref{eq:Mpi1pi2approx}). Note that while this exercise was done here with a simple analytical forward model as an illustration of the procedure, it can be easily repeated with a full numerical forward model.

\subsection{Order of magnitude analysis}
\label{sec:orderofmagnitude}

\begin{figure}[t]
\centering
\includegraphics[width=.8\textwidth]{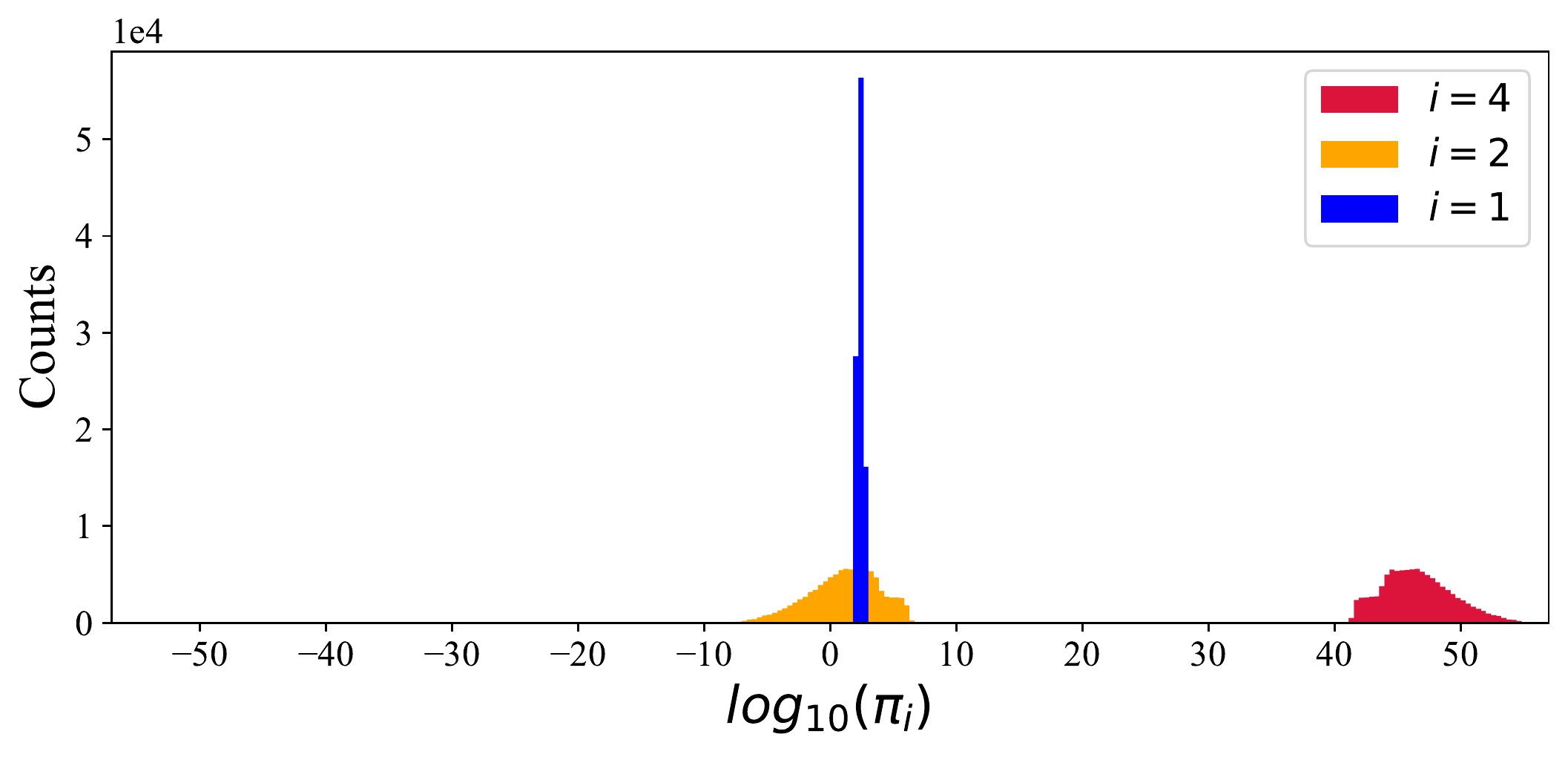}
\caption{Distributions of the (base ten logarithms of the) dimensionless quantities $\pi'_1$, $\pi_2$ and $\pi'_4$ in the data set of \cite{Marquez2018}.
\label{fig:pies}}
\end{figure}

Dimensional analysis does provide a rule of thumb for determining whether a given governing parameter is relevant or not: ``if the dimensionless parameter is either very small or very large compared to unity, it may be assumed to be not essential, and the function $f$ can be assumed to be constant (or, in general, when there are several dimensionless parameters, a function of one fewer arguments)" \citep{Barenblatt1996}. Let us check if this is the case in our example. Figure~\ref{fig:pies} shows distributions of (base ten logarithms of) the dimensionless quantities $\pi'_1$, $\pi_2$ and $\pi'_4$ in the data set of \cite{Marquez2018} (recall that $\pi_3$ is constant throughout the data set, so adding its distribution to Fig.~\ref{fig:pies} would only add a trivial delta function near the origin).

Figure~\ref{fig:pies} shows that while the distributions of $\pi'_1$ and $\pi_2$ are in the proximity of $10^0=1$, the distribution of $\pi'_4$ is shifted by almost 50 orders of magnitude. The expectations from dimensional analysis would then suggest that $\pi'_1$ and $\pi_2$ are essential variables while $\pi'_4$ is a non-essential variable. This conclusion is in agreement with the discussion from the previous subsection. 

\subsection{Direct tests of the function $f$}

While the relevancy tests in the previous two subsections relied on approximations, heuristics or intuition from dimensional analysis, in this subsection we shall perform a direct quantitative test of the behavior of the function $f$ with respect to its arguments. In particular, we shall be changing one Pi group at a time, while keeping the other Pi groups constant, and then we shall be checking whether the variation of a single Pi group leads to a change in $f(\pi_k)$ or not. We imagine that in practice this test would be done with a full numerical forward radiative transfer model, but for the purpose of this paper, it is sufficient to illustrate the basic idea with the numerical forward model (\ref{eq:simpleformula}) used to generate the data set of \cite{Marquez2018}.

\begin{figure*}[t]
\centering
\includegraphics[width=.95\textwidth]{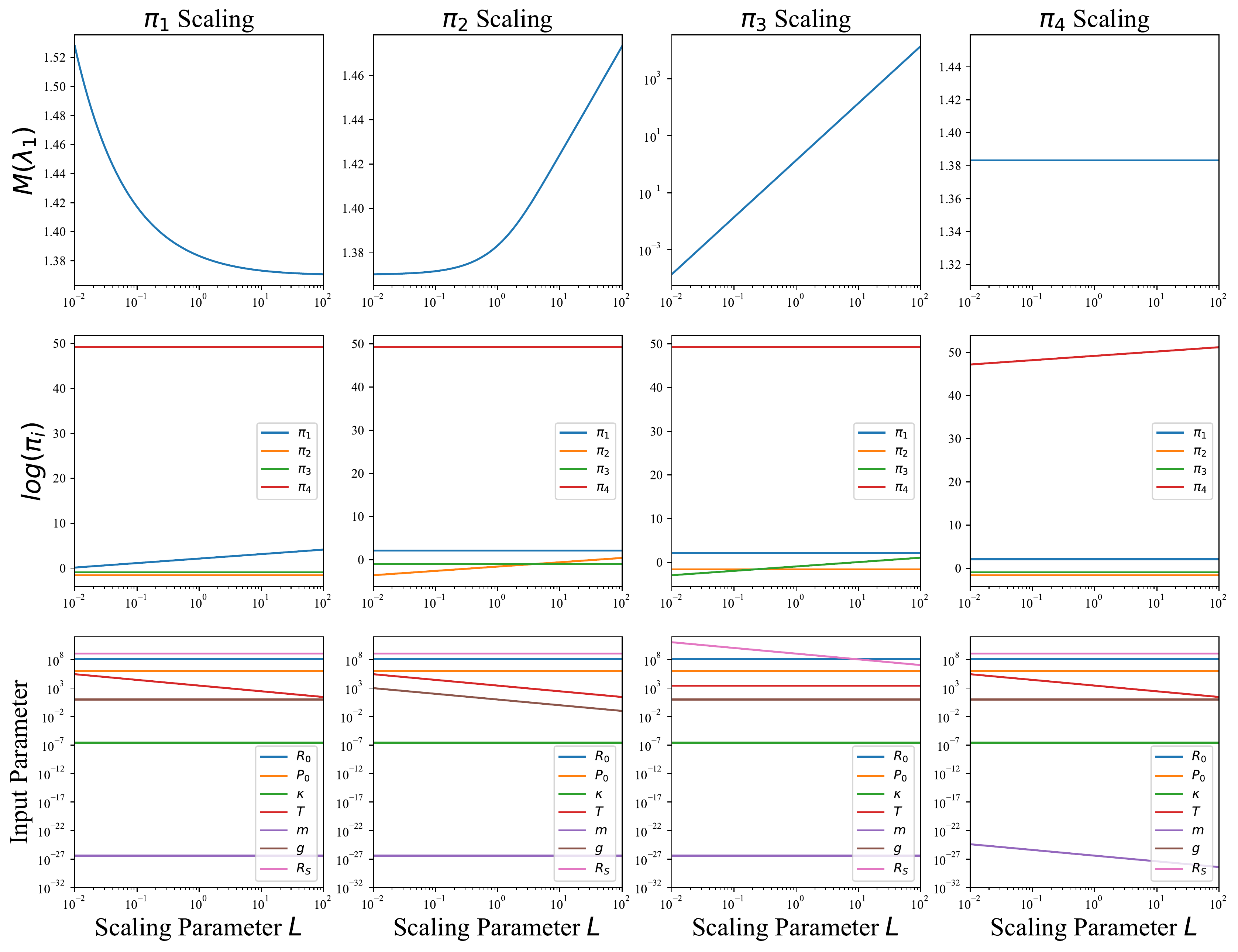}
\caption{Testing the dependence of the modulation $M(\lambda_1)$ on each individual $\pi_k$ parameter. The plots in the bottom row show the scaling of the input parameters from Table~\ref{tab:variables} which is needed to vary one of the $\pi$'s but not the others, as illustrated by the plots in the middle row. The plots in the top row show the resulting variation in $M(\lambda_1)$.
\label{fig:scaling}}
\end{figure*}

The results from the exercise are shown in Figure~\ref{fig:scaling}. The bottom panels show the required variation of the input variables from Table~\ref{tab:variables} which triggers a variation in only one of the $\pi_k$'s while keeping the others constant, as illustrated by the plots in the middle row. The exact scaling relations are as follows:
\begin{subequations}
\begin{eqnarray}
T\to T/L ~~&\Longrightarrow&~~ \pi'_1 \to L \pi'_1, \quad \pi_2, \pi_3 {\rm\ and\ } \pi'_4 {\rm \ remain\ constant}; \label{eq:pi1scaling} \\
T\to T/L;\quad g\to g/L ~~&\Longrightarrow&~~ \pi_2 \to L \pi_2, \quad \pi'_1, \pi_3 {\rm\ and\ } \pi'_4 {\rm \ remain\ constant};  \label{eq:pi2scaling} \\
R_S\to R_S/L ~~&\Longrightarrow&~~ \pi_3 \to L \pi_3, \quad \pi'_1, \pi_2 {\rm\ and\ } \pi'_4 {\rm \ remain\ constant}; \label{eq:pi3scaling} \\
T\to T/L;\quad m\to m/L ~~&\Longrightarrow&~~ \pi'_4 \to L \pi'_4,  \quad \pi'_1, \pi_2 {\rm\ and\ } \pi_3 {\rm \ remain\ constant}.  \label{eq:pi4scaling}
\end{eqnarray}
\label{eq:piscaling}
\end{subequations}
The top panels in Figure~\ref{fig:scaling} reveal that $M(\lambda_1)$ depends on $\pi'_1$, $\pi_2$ and $\pi_3$, but not on $\pi'_4$. This result is in agreement with our conclusions from Sections \ref{sec:comparison} and \ref{sec:orderofmagnitude}, only this time it is made on a firm quantitative footing. Note that while we have established the dependence of $M(\lambda_1)$ on $\pi'_1$, $\pi_2$ and $\pi_3$, we still do not know its exact functional form --- this task is postponed for Section~\ref{sec:regression}.

\subsection{The complete set of guaranteed and accidental degeneracies}

\begin{figure}[t]
\centering
\includegraphics[width=.9\textwidth]{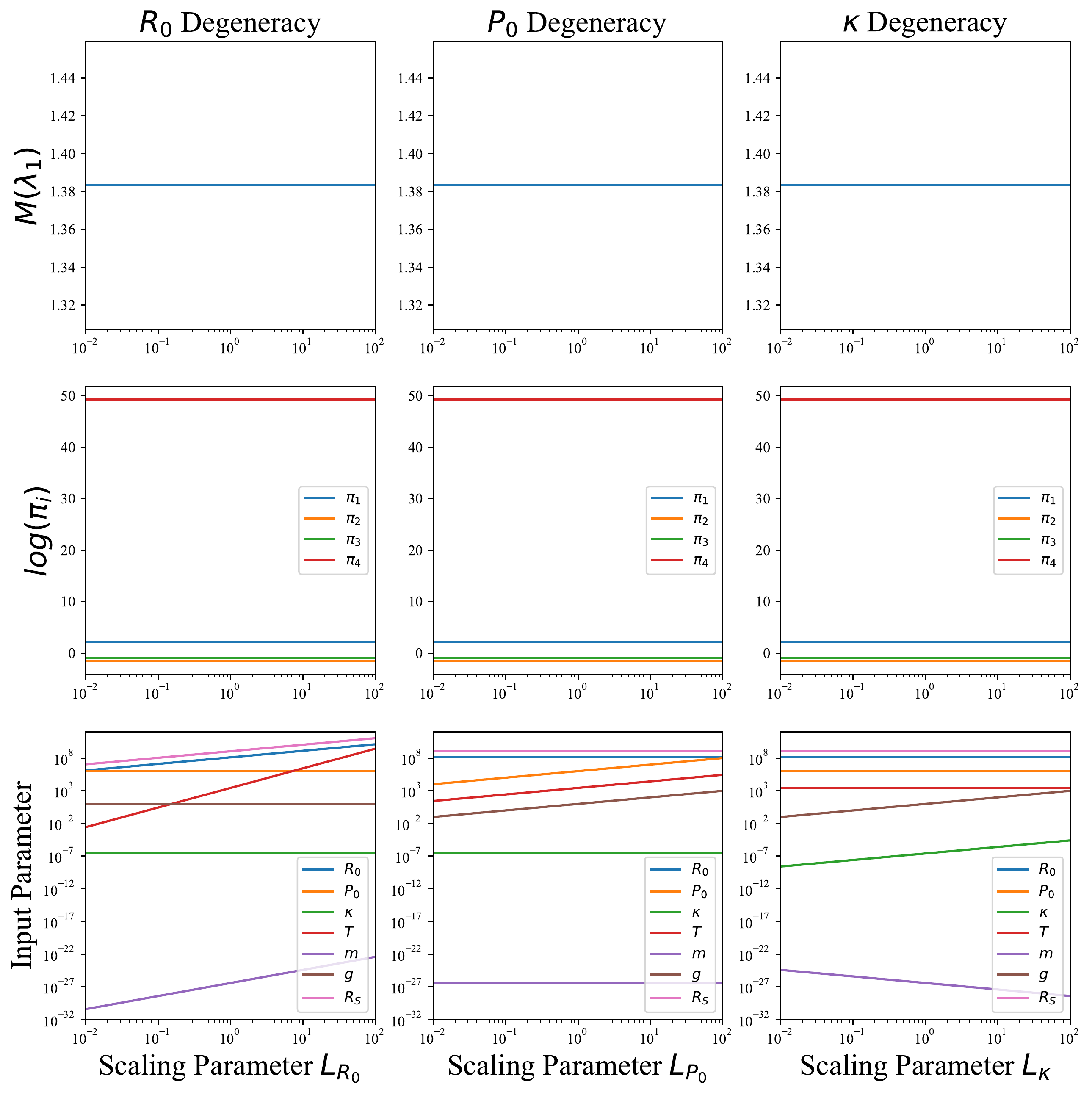}
\caption{The same as Figure~\ref{fig:scaling}, but this time varying the input parameters according to eqs.~(\ref{eq:Lscalings}). In the lower left, lower middle and lower right panels the parameters are varied according to eqs.~(\ref{eq:LR0}), (\ref{eq:LP0})  and (\ref{eq:Lkappa}), respectively. The plots in the middle and top rows confirm that these transformations are indeed true degeneracies.
\label{fig:degeneracies}}
\end{figure}

The analysis in the previous subsection revealed that $\pi'_4$ is a cyclic variable, and its value does not affect the observed transmission spectrum. This leads to an additional, ``accidental", degeneracy, illustrated by the plots in the rightmost column in Figure~\ref{fig:scaling}. Using eq.~(\ref{eq:pi4scaling}), we can express this degeneracy as
\begin{subequations}
\begin{eqnarray}
T ~~&\to&~~ L_{T} \times T, \\
m ~~&\to&~~ L_{T} \times m
\end{eqnarray}
\label{eq:accidental}
\end{subequations}
in terms of a new scaling parameter $L_T$. This degeneracy implies that if the mean molecular mass is unknown a priori, there would be difficulties in extracting the precise value of the temperature.

Recall that in Section~\ref{sec:guaranteed} we already derived a family of guaranteed degeneracies (\ref{eq:degeneracyscalings}) among the seven input atmospheric variables. For completeness, Figure~\ref{fig:degeneracies} illustrates those earlier discussions in the same format as Figure~\ref{fig:scaling}. Now we can combine the results of eqs.~(\ref{eq:degeneracyscalings}) and (\ref{eq:accidental}) to arrive at the most general family of degeneracies in $M(\lambda)$
\begin{subequations}
\begin{eqnarray}
R_0 ~~&\to&~~ L_{R_0} \times R_0 \\
P_0 ~~&\to&~~ L_{P_0} \times P_0 \\
\kappa ~~&\to&~~ L_{\kappa} \times \kappa \\
T ~~&\to&~~ L^3_{R_0} L_{P_0} L_{T} \times T \\
m ~~&\to&~~ L^2_{R_0} L_{\kappa}^{-1} L_{T}\times m \\
g ~~&\to&~~ L_{P_0} L_{\kappa}  \times g \\
R_S ~~&\to&~~ L_{R_0} \times R_S 
\end{eqnarray}
\label{eq:completedegeneracy}
\end{subequations}
which include both the degeneracies guaranteed by dimensional analysis alone and the accidental degeneracy arising due to the specific functional form of $f$.

Equation (\ref{eq:completedegeneracy}) is one of the main results of this paper. Being completely general, it should encapsulate all existing discussions on degeneracies in the literature; i.e., one should be able to reproduce any\footnote{Note that additional degeneracies not captured by eq.~(\ref{eq:completedegeneracy}) may arise when one tries to disentange the contributions of individual minor gases and/or clouds to the overall opacity of the atmosphere $\kappa$.} previous correctly identified degeneracy as a special case of (\ref{eq:completedegeneracy}). For illustration, in Appendix~\ref{sec:appendix} we have compiled a list of possible degeneracies involving two or three atmospheric parameters, summarized pictorially in Figure~\ref{fig:diagram}. Note that, according to Figure~\ref{fig:diagram}, the specific gravity $g$ plays a central role in the existing degeneracies. This suggests that pinning down the value of $g$ to high accuracy by means of independent observations would eliminate a large number of degeneracies and thus significantly decrease the uncertainties on the remaining parameters. 

\section{Symbolic Regression}
\label{sec:regression}

Symbolic regression is a somewhat underused, yet very interpretable machine learning algorithm for modelling a data set with analytic expressions by searching over a suitable space of functions. Since the numerical forward models of radiative transfer are quite complicated, the problem of finding valid and easily interpretable analytical expressions reproducing the results from the numerical models is well motivated.

In this section we shall fit a symbolic regression to transit spectroscopy data. For this proof of concept exercise, we shall use data generated by the analytical forward model (\ref{eq:simpleformula}) behind the benchmark data set of \cite{Marquez2018}. Since the performance of symbolic regressions is known to deteriorate as the dimensionality of the data increases, we shall take advantage of our earlier dimensional analysis which allowed us to reduce the number of independent inputs from 7 to 4, namely $\pi_1$, $\pi_2$, $\pi_3$ and $\pi_4$.\footnote{To simplify the notation, from now on in this section we shall omit the primes on $\pi'_1$ and $\pi'_4$.} To be specific, we shall try to learn the dimensionless function
\begin{equation}
f_\text{true}(\pi_1,\pi_2,\pi_3,\pi_4) \equiv \frac{R_T(\lambda)}{R_S}-1 = \frac{1}{\pi_1} \left[\gamma_E  + \ln \left(\pi_2\sqrt{2\pi \pi_1}\right)\right].
\label{eq:ftrue}
\end{equation}
To do this, we shall make use of the recently released PySR software package \citep{Cranmer:2020wew}. It models the data set with a graph neural network before applying symbolic regression to fit different internal parts of the learned model that operate on reduced dimension representations. We shall not attempt any hyperparameter optimization and instead we will use the default configuration in the PySR distribution. 

\begin{figure}[t]
\centering
\includegraphics[width=.95\textwidth]{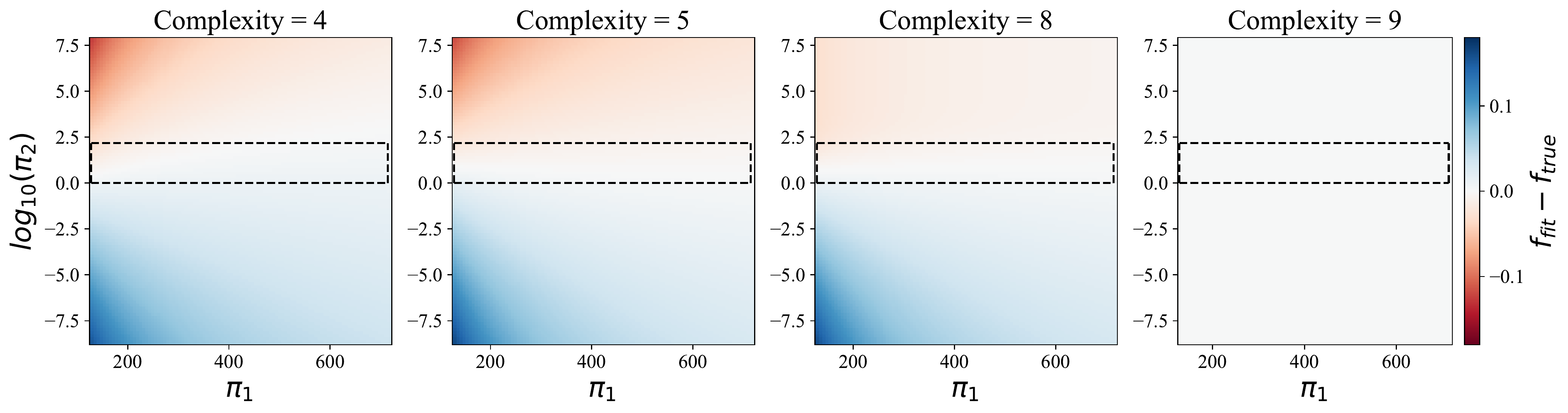}
\caption{Heatmaps of the differences between the $f_\text{fit}^{(C)}$ functions (\ref{eq:fits}) obtained by symbolic regression and the target function $f_\text{true}$ (\ref{eq:ftrue}). The rectangular box marked with a dashed line delineates the domain of values on which the symbolic regression was trained.
\label{fig:errors}}
\end{figure}

We generate training data of 1000 samples as follows. We sample $\pi_1$ within its full range in the benchmark data set of \cite{Marquez2018}, namely from 125 to 714. As shown in Figure~\ref{fig:pies}, the values of $\pi_2$ in the data set span many orders of magnitude, thus we choose to sample $\pi_2$ only within a restricted range from $e^0=1$ to $e^5=148$. The dashed rectangle in Figure~\ref{fig:errors} shows the resulting region in the $(\pi_1,\log_{10}\pi_2)$ parameter space where the training data was generated. In addition, we also generate values for $\pi_3$ and $\pi_4$ (sampled uniformly between -1 and 1), despite the fact that the function $f_\text{true}$ does not explicitly depend on them --- instead, we let the symbolic regression figure out on its own that those are cyclic variables. In summary, the input to the symbolic regression is a set of 1000 instances of the form
$$
(\pi_1,\pi_2,\pi_3,\pi_4,f_\text{true}).
$$

The result from a PySR run is a set of functions $f_\text{fit}^{(C)}$ of increasing complexity $C$ (defined as the number of leaf nodes in the binary tree representing the analytical expression for $f_\text{fit}$). The result from one typical run for the fitted functions and the corresponding mean squared error (MSE) is
\begin{subequations}
\begin{align}
f^{(4)}_\text{fit} &=
\frac{0.456}{\sqrt{|\pi_1|}} ,
&MSE = 8.314751 \times 10^{-5},
\\ 
f^{(5)}_\text{fit} &=
\frac{1.9413}{0.3009\, \pi_1}
=\frac{6.452}{\pi_1} ,
&MSE = 6.008056 \times 10^{-5}	,
\\ 
f^{(8)}_\text{fit} &=
\frac{\ln(|\pi_2+2.88\,\pi_1|)}{\pi_1}, 
&MSE = 5.702897\times 10^{-5},
\\ 
f^{(9)}_\text{fit} &=
\frac{\ln(4.4645\, |\pi_2|\,\sqrt{|\pi_1|})}{\pi_1},
&MSE = 1.003324 \times 10^{-15}.
\label{eq:ffit9}
\end{align}
\label{eq:fits}
\end{subequations}

We see that the symbolic regression was able to correctly determine that $\pi_3$ and $\pi_4$ are irrelevant variables --- none of the proposed formulas involve $\pi_3$ or $\pi_4$. Of course, when we demand a low complexity like $C=4$ or $C=5$, the program is forced to make a choice between $\pi_1$ and $\pi_2$ to derive a good fit, because using both would exceed the set complexity threshold.

Figure~\ref{fig:errors} compares quantitatively the results from the fit (\ref{eq:fits}) to the true function (\ref{eq:ftrue}). We show heatmaps of the difference $(f^{(C)}_\text{fit}-f_\text{true})$ in the $(\pi_1,\log_{10}\pi_2)$ parameter space. The rightmost panel in the figure demonstrates that at complexity level 9 the symbolic regression was able to nail down the target function exactly, to within the machine precision of $10^{-15}$. Indeed, the original expression (\ref{eq:ftrue}) can be equivalently rewritten as
\begin{equation}
f_\text{true}(\pi_1,\pi_2,\pi_3,\pi_4) 
= \frac{\ln\left[\left(e^{\gamma_E}\sqrt{2\pi}\right)\pi_2\sqrt{\pi_1}\right]}{\pi_1}
= \frac{\ln\left(4.4645\pi_2\sqrt{\pi_1}\right)}{\pi_1},
\label{eq:ftruealternative}
\end{equation}
in perfect agreement with (\ref{eq:ffit9}). The other three panels in Figure~\ref{fig:errors} show that, as expected, at lower complexities the symbolic regression is unable to fit the data perfectly, although the fit is good within the training region identified with the dashed rectangular box.

\section{Conclusions and Outlook}

Observation of planetary transits at different wavelengths is a widely used technique to extract information about the structure and the composition of the atmosphere of an exoplanet. As shown in Figure~\ref{fig:flowchart}, the methods used to constrain the atmospheric parameters can be generally divided into two groups: 
\begin{enumerate}
\item Computational studies based on forward radiative transfer models and numerical inversion techniques, which increasingly utilize different novel statistical and machine learning methods to improve the accuracy, the precision, and the speed of the performed retrievals.
\item Analytical investigations, which, provided with a set of simplifying assumptions, derive an analytical expression that allows for a better understanding of the underlying physics.
\end{enumerate}
In this paper we demonstrate a novel approach to the problem (the central path in Figure~\ref{fig:flowchart}) which leverages the advantages of these two approaches by harvesting the data generated by the complex and detailed forward models while preserving the ability of analytical expressions to provide physical insight into the problem.

First, we perform a formal dimensional analysis on the modulated stellar spectrum $M(\lambda)$ during planetary transit, as a function of the seven input atmospheric parameters listed in Table~\ref{tab:variables}: 
\begin{itemize}
\item We show that $M(\lambda)$ depends on only three dimensionless groups: 
$$\pi'_1 = \frac{R_0mg}{k_BT}, \qquad \pi_2 = \frac{P_0\kappa}{g},\qquad \pi_3 = \frac{R_0}{R_S}.$$
\item We mathematically demonstrate that the transit spectrum suffers from a number of degeneracies summarized in eq.~(\ref{eq:completedegeneracy}). The simplest two- and three-level degeneracies are collected in Appendix~\ref{sec:appendix} and illustrated in Fig.~\ref{fig:diagram}. The higher-level degeneracies which involve more than 3 parameters can be obtained from the equation set (\ref{eq:completedegeneracy}). Some of the degeneracies have been previously noted in the literature, however, we base our study on mathematically rigorous grounds which allows us to find {\em all} theoretically possible degeneracies between the relevant planetary parameters.
\item We discuss methods for lifting the degeneracies by using additional observations or theoretical/model constraints on the stellar radius, gravity acceleration of the planet and/or the mean molecular mass of the planet atmosphere.
\item The dimensionality reduction of the input parameter space achieved by our dimensional analysis helps greatly in optimizing the performance of the symbolic regression algorithm used in the second half of the paper.
\end{itemize}

In the second part of the paper, as a proof-of-concept we demonstrate the use of symbolic regression to derive analytical expressions describing the relationship between the modulated stellar spectrum $M(\lambda)$ and the input atmospheric parameters. The regression takes advantage of the dimensionality reduction accomplished by our dimensional analysis, and is performed on a synthetic data set generated with the analytical model of \cite{Heng2017}, in analogy to \cite{Marquez2018}. The algorithm correctly identifies both the relevant and the irrelevant variables, and provides analytic approximations of increasing complexity $C$, eventually reproducing the correct expression perfectly at $C=9$ (to within the machine precision). 

The analysis presented in this paper can be extended in several interesting directions.
\begin{itemize}
    \item Expand the dimensional analysis by including additional variables which account for further details like temperature gradients, planet rotation, etc., which would introduce new possible degeneracies.
    \item Perform the symbolic regression on a suitable data set generated by a full forward radiative transfer model scanning all relevant parameters ($\pi$ groups).
    \item Explore other modern techniques for symbolic regression which are adaptable to high dimensional data.
\end{itemize}

\begin{acknowledgments}
We thank K.~Kong (University of Kansas) for collaboration in the early stages of this work. This work was supported in part by the United States Department of Energy under Grant No. DESC0022148.
\end{acknowledgments}


\software{
{\tt jupyter} \citep{Kluyver2016},
{\tt matplotlib} \citep{Hunter:2007ouj},
{\tt numpy} \citep{vanderWalt2011},
{\tt pysr} \citep{Cranmer:2020wew},
{\tt scipy} \citep{Scipy2020}.
}


\appendix

\section{List of Two-Level and Three-Level Degeneracies}
\label{sec:appendix}

\begin{figure}[t]
\centering
\includegraphics[width=.55\textwidth]{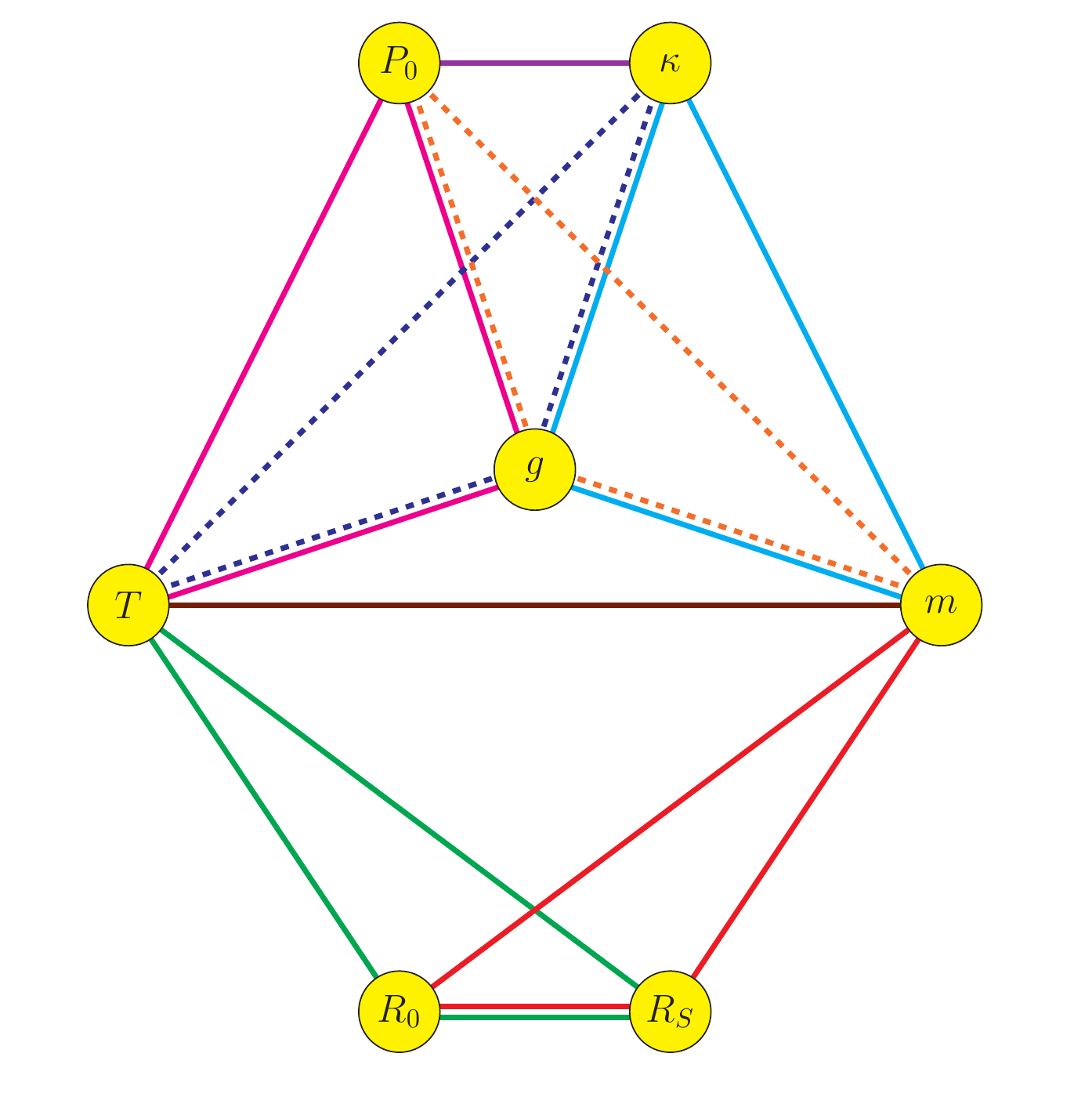}
\caption{A pictorial representation of the two-level (line segments) and three-level (triangles) degeneracies discussed in the text. Each color corresponds to a different set of degenerate parameters, i.e., parameters which are not uniquely fixed by the experimental observation (two triangles were drawn with dashed lines for easier identification).
\label{fig:diagram}}
\end{figure}

In this appendix we list the simplest degeneracies implied by eq.~(\ref{eq:completedegeneracy}), i.e., the degeneracies involving the fewest number of atmospheric parameters (see Fig.~\ref{fig:diagram} for a pictorial summary).

Let us start with the two-level degeneracies:
\begin{itemize}
    \item {\em Pressure-opacity degeneracy} (deep purple line segment in Fig.~\ref{fig:diagram}).  \cite{Heng2017} pointed out a degeneracy between the pressure $P_0$ and the mass mixing ratio which is incorporated into the calculation of $\kappa$. We can reproduce this degeneracy from eq.~(\ref{eq:completedegeneracy}) by taking $L_{R_0}=1$ and $L_\kappa=L_T=L_{P_0}^{-1}\equiv L^{-1}$, which results in
\begin{equation}
 \quad P_0 \to L \times P_0, \quad \kappa \to L^{-1} \times \kappa.
\label{eq:PK}
\end{equation}
    \item {\em Temperature-mean molecular mass degeneracy} (brown line segment in Fig.~\ref{fig:diagram}).  This is the accidental degeneracy shown in our previous result (\ref{eq:accidental}):
\begin{equation}
 \quad T \to L \times T, \quad m \to L \times m.
 \label{eq:TM}
\end{equation}
\end{itemize}

There are also six three-level degeneracies:
\begin{itemize}
\item {\em Pressure-temperature-gravity degeneracy} (magenta solid triangle in Fig.~\ref{fig:diagram}). This is one of the guaranteed degeneracies shown in our previous result (\ref{eq:LP0}):
\begin{equation}
P_0 \to L \times P_0,\quad T \to L \times T, \quad g \to L \times g.
 \label{eq:PTG}
\end{equation}    
\item {\em Opacity-temperature-gravity degeneracy} (blue dashed triangle in Fig.~\ref{fig:diagram}).
By choosing $L_{\kappa}=L_T\equiv L$ and $L_{R_0}=L_{P_0}=1$ in eq.~(\ref{eq:completedegeneracy}), we obtain 
\begin{equation}
\kappa \to L \times \kappa,\quad T \to L \times T, \quad g \to L \times g.
 \label{eq:OTG}
\end{equation}        
\item {\em Pressure-mass-gravity degeneracy} 
(orange dashed triangle in Fig.~\ref{fig:diagram}). By choosing $L_{P_0}=L_T^{-1}\equiv L$ and $L_{R_0}=L_{\kappa}=1$ in eq.~(\ref{eq:completedegeneracy}), we obtain 
\begin{equation}
P_0 \to L \times P_0,\quad m \to L^{-1} \times m, \quad g \to L \times g.
 \label{eq:PMG}
\end{equation}     
\item {\em Opacity-mass-gravity degeneracy} (cyan solid triangle in Fig.~\ref{fig:diagram}). This is another one of the guaranteed degeneracies discussed earlier in (\ref{eq:Lkappa}):
\begin{equation}
\kappa \to L \times \kappa,\quad m \to L^{-1} \times m, \quad g \to L \times g.
 \label{eq:KMG}
\end{equation}        
\item {\em Planet radius - star radius - temperature degeneracy} (green solid triangle in Fig.~\ref{fig:diagram}). By choosing $L_{R_0}=L_T^{-1/2}\equiv L$ and $L_{P_0}=L_\kappa=1$ in eq.~(\ref{eq:completedegeneracy}), we obtain 
\begin{equation}
R_0 \to L \times R_0,\quad T \to L \times T, \quad R_S \to L \times R_S.
 \label{eq:RTR}
\end{equation}        
\item {\em Planet radius - star radius - mass degeneracy} (red solid triangle in Fig.~\ref{fig:diagram}). 
By choosing $L_{R_0}=L_T^{-1/3}\equiv L$ and $L_{P_0}=L_\kappa=1$ in eq.~(\ref{eq:completedegeneracy}), we obtain 
\begin{equation}
R_0 \to L \times R_0,\quad m \to L^{-1} \times m, \quad R_S \to L \times R_S.
 \label{eq:RmR}
\end{equation}        
\end{itemize}

These simple degeneracies can be combined together to form degeneracies involving more than three atmospheric parameters.


\bibliography{symreg}{}
\bibliographystyle{aasjournal}



\end{document}